\documentclass[a4paper]{jpconf}
\usepackage{graphicx}
\usepackage{xcolor}
\begin{document}
\title{3-Dimensional QCD Phase Diagrams \\for Strange Matter}

\author{V. Dexheimer, K. Aryal, C. Constantinou, J. Peterson}
\address{Department of Physics, Kent State University, Kent, OH 44243 USA}
\author{R. L. S. Farias}
\address{Departamento de F\'{\i}sica, Universidade Federal de Santa Maria,
97105-900 Santa Maria, RS, Brazil}

\ead{vdexheim@kent.edu}

\begin{abstract}

In this work, we examine in detail the difference between constraining the electric charge fraction and isospin fraction when calculating the deconfinement phase transition in the presence of net strangeness. We present relations among charge and isospin fractions and the corresponding  chemical potentials and draw 3-dimensional QCD phase diagrams for matter out of weak  equilibrium. Finally, we briefly discuss how our results can be applied to comparisons of matter created in heavy ion collisions and binary neutron star mergers.

\end{abstract}

\section{Introduction and Formalism}
\vspace{1mm}

When presenting phase diagrams for quantum chromodynamics (QCD), most studies only look in two dimensions at once; usually, temperature and either baryon chemical potential or isospin chemical potential. In the former case, the charged chemical potential is set to minus the electron chemical potential $\mu_Q=-\mu_e$ (to describe cold neutron stars \cite{Dexheimer:2008ax}) or the isospin chemical potential is set to zero $\mu_I=0$ (to describe very energetic relativistic heavy-ion collisions~\cite{Stephanov:2007fk}). In the latter case, the baryon chemical potential is set to zero $\mu_B=0$~\cite{PhysRevD.97.054514}. See Refs.~\cite{Hempel:2013tfa,Roark:2018uls} for details on changes that appear in 2-dimensional phase diagrams in which either the charge fraction is fixed at $Y_Q=0.3$ or the lepton fraction at $Y_L=0.4$.

Several studies have compared extreme astrophysical environments (e.g. binary neutron star mergers, core collapse supernovae, and proto-neutron stars) to heavy ion collisions (HIC) \cite{Hanauske:2017oxo,Bastian:2018wfl}, motivated by the high (compared to Fermi) temperatures generated in all cases. However, recent simulations \cite{Most:2019onn} suggest that neutron star mergers are incapable of reaching the high charge fractions typically associated with HICs and young/hot neutron stars. Acknowledging this variety of conditions, we explore the consequences of varying the charge and isospin fractions and use either of these quantities as a third axis in the QCD phase diagram, focusing on changes to the first order phase transition coexistence line between the hadronic and the quark phases. 

Complimentary to what was done recently in Ref.~\cite{Aryal:2020ocm}, where 3-dimensional phase diagrams were shown for non-strange matter, here we introduce 3-dimensional phase diagrams with net strangeness, a case consistent with astrophysical processes occurring over timescales much longer than those of weak interactions. We do not impose weak  equilibrium, but vary freely the charged or isospin chemical potential of the system (depending on the respective chosen constrain, charge or isospin fraction) within physically meaningful values.

For this purpose, we make use of the Chiral Mean Field (CMF) formalism. It is  a non-linear realization of the SU(3) linear sigma model for hadronic matter, which in its present version also contains contributions from quarks  \cite{Dexheimer:2008ax} \footnote{Note that an alternative version of the CMF model includes in addition the chiral partners of the baryons and gives the baryons a finite size \cite{Steinheimer:2011ea, Motornenko:2019arp}.}. As a result, this formalism reproduces chirally symmetric matter with quarks at large densities and/or temperatures, the phase transition being of first order, except when it becomes a crossover at large temperatures. The hadronic sector of the model was fitted to various physical properties including particle vacuum masses, decay constants, nuclear saturation properties, symmetry energy, and reasonable values for hyperon potentials. The quark sector of the model reproduces expectations of the phase diagram from heavy-ion collisions and lattice QCD. In addition, at high densities this formalism was shown to be in good agreement with perturbative QCD for the conditions expected to be found in different astrophysical scenarios \cite{Roark:2018uls}.

\section{Results and Discussion}
\vspace{1mm}

\begin{figure*}[t!]
\centering
\begin{minipage}{17.96pc}
\includegraphics[width=\textwidth]{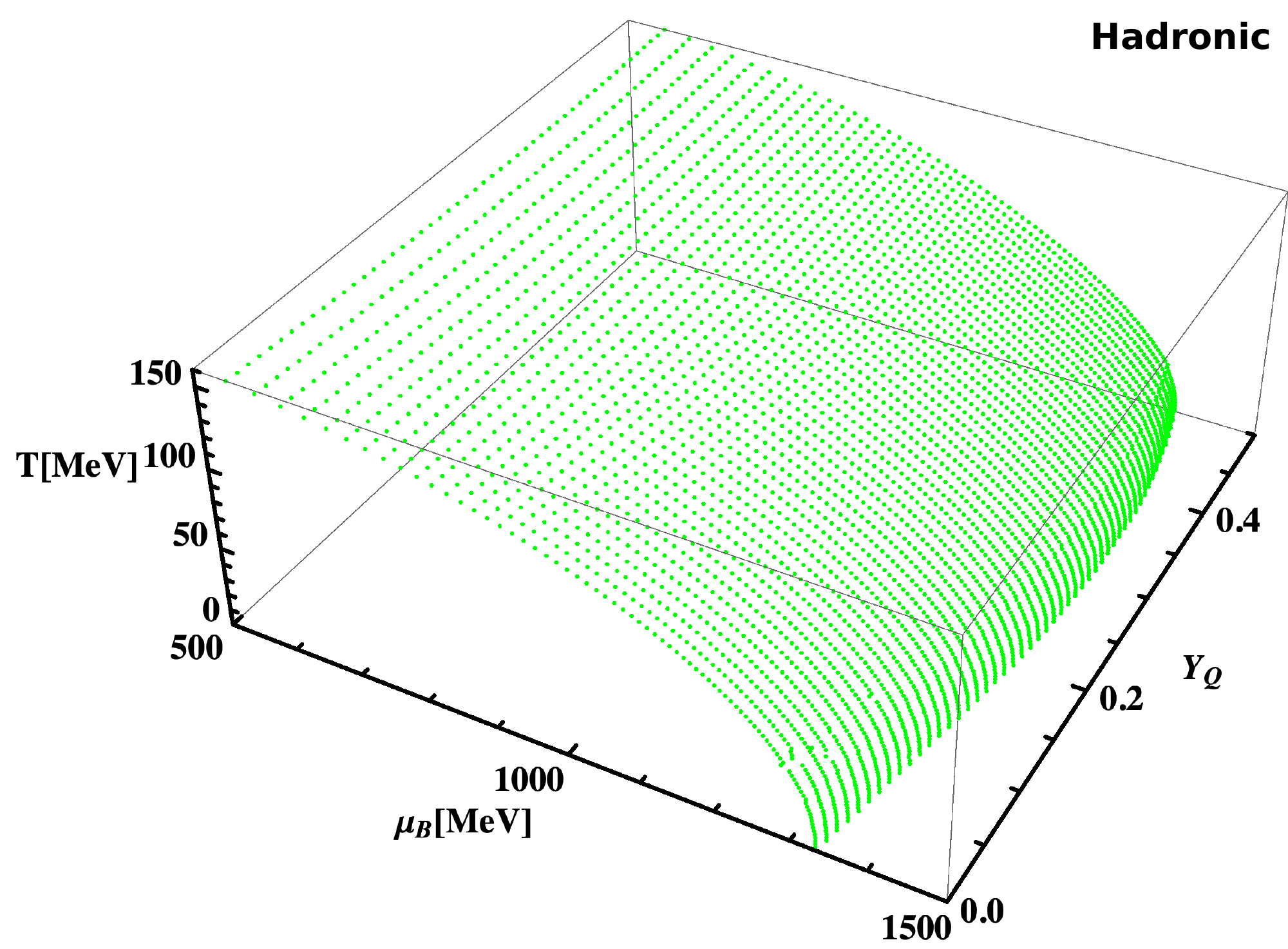}
\end{minipage}\hspace{2pc}%
\begin{minipage}{17.96pc}
 \includegraphics[width=\textwidth]{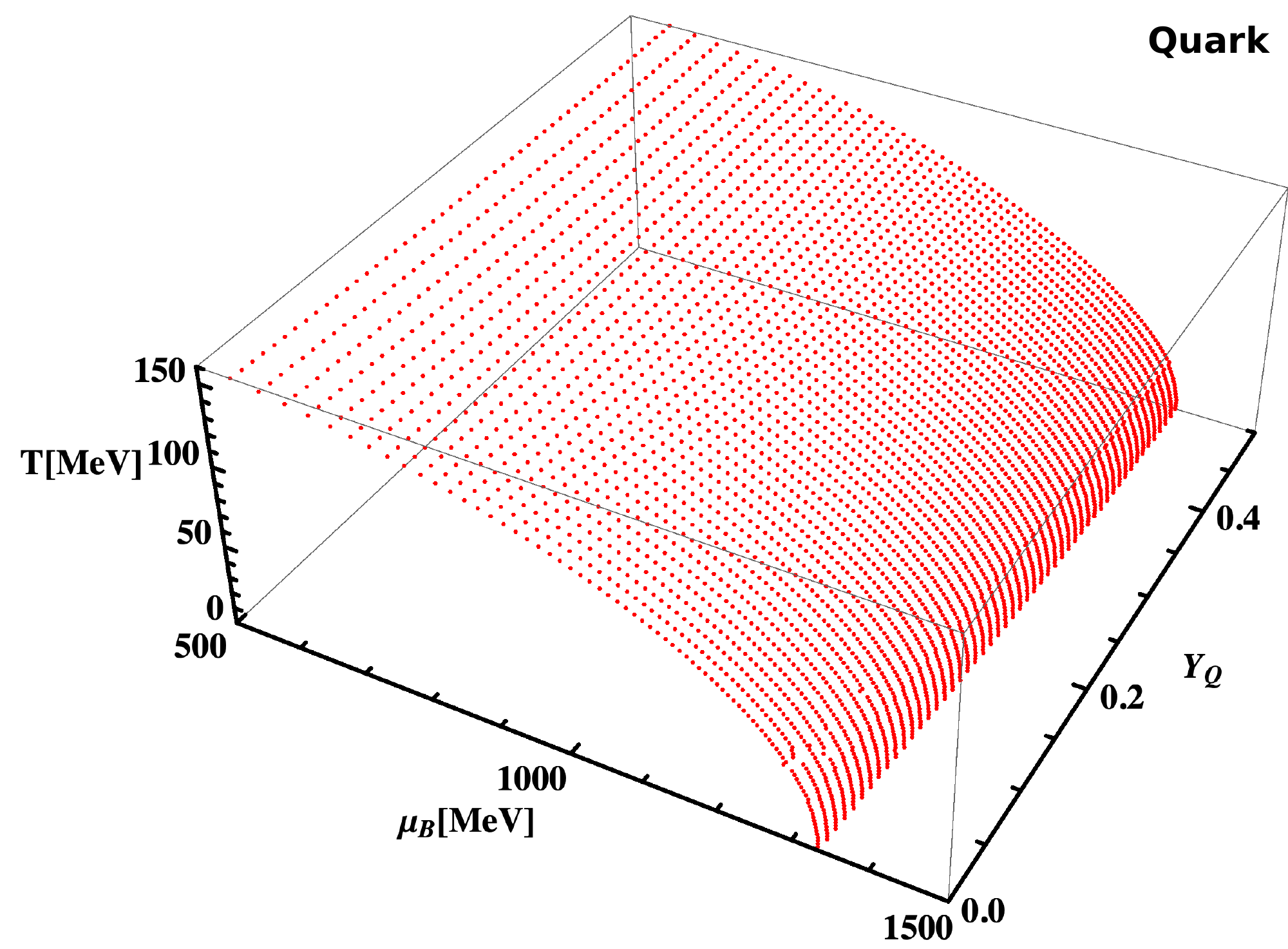}
\end{minipage}\hspace{2pc}%
\begin{minipage}{17.96pc}
\includegraphics[width=\textwidth]{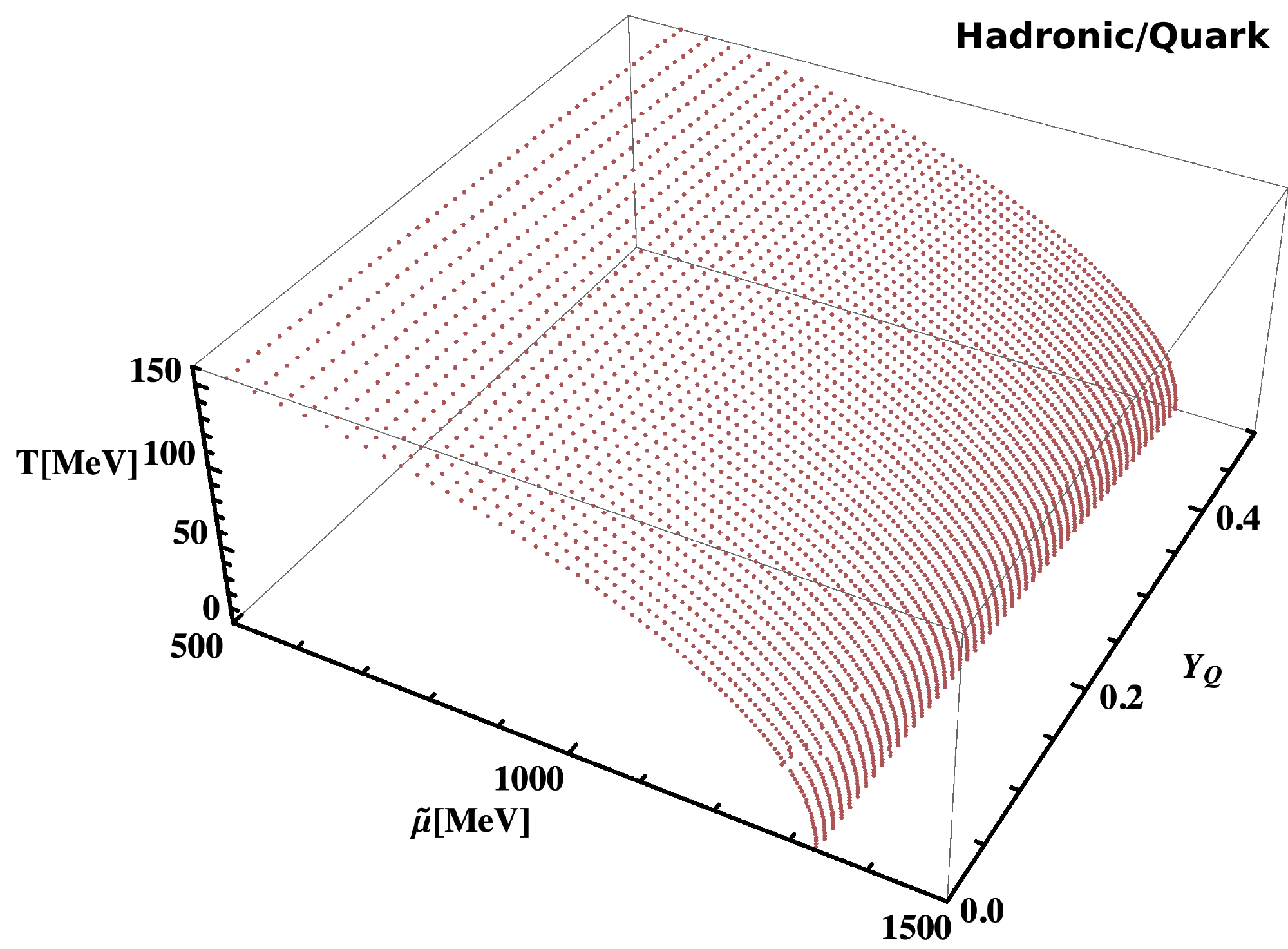}
\end{minipage} 
  \caption {Top panels: the temperature T vs$.$ baryon chemical potential $\mu_B$ vs$.$ charge fraction $Y_Q$ phase diagram on the hadronic side of the deconfinement phase transition (left panel) and on the quark side (right panel). Bottom panel: the temperature T vs$.$ free energy $\widetilde{\mu}$ vs$.$ charge fraction $Y_Q$ phase diagram either on the hadronic or quark side of the deconfinement phase transition. All curves were calculated varying the charge fraction between $Y_Q=0$ and  $0.5$.}
\label{fig:fig}
\end{figure*} 

Fig.~1 is comprised of different 3-dimensional phase diagrams built using the CMF model. The top two panels show the temperature vs. baryon chemical potential vs. charge fraction on either the hadronic side of the deconfinement phase transition (left panel) or on the quark side (right panel). The differences between the two are due to the extra constraint of conserved charge fraction, which is varied between $0$ and $0.5$. For this work, we choose to stop all our phase diagrams at a temperature $T=160$ MeV (around the critical temperature for the first order phase transition), as a more thorough examination of this feature is beyond the scope of the current discussion. 

The charge fraction is defined as the total electric charge of a system 
normalized by the total particle number of the system. For a baryon/quark mixture, the former quantity can be expressed as a sum (over all baryon and quark species) of the individual electric charge - number density products. Similarly, the latter is a sum of all baryon number - number density products
\begin{eqnarray}
Y_Q = \frac{Q}{B} = \frac{\sum_i Q_i \ n_i} {\sum_i Q_{B,i} \ n_i} .
\label{6h}
\end{eqnarray}
In particular, having $Y_Q = 0$ means that there is no net charge in the system, even though the presence of charged particles is not prohibited insofar as the sum of their charges is zero. Having $Y_Q = 0.5$ corresponds to the situation in which the total number of baryons of the system is twice as large as its net charge (in units of the electron charge). When charge neutrality is imposed, for example to model neutron stars, the (hadronic and quark) charge fraction must be equal to the lepton fraction. In this work, we are not showing contributions from leptons, as they are not in weak equilibrium with hadrons and quarks for the large temperatures in which we are interested.

In the bottom panel of Fig.~1, the temperature vs. free energy $\widetilde{\mu}$ vs. charge fraction phase diagram is shown. The Gibbs free energy per baryon (shortened here to free energy) is by the requirement of chemical stability the same on the hadronic and quark sides of the deconfinement phase transition. In our case when, besides baryon number, electric charge or isospin are conserved, it is
\begin{eqnarray}
\widetilde{\mu} &=& \mu_B + Y_Q\mu_Q ,\label{ve2Q}
\end{eqnarray}
or
\begin{eqnarray}
\widetilde{\mu} &= &\mu_B + (Y_I + {{1}/{2}} ) \mu_I ,
\label{ve2}
\end{eqnarray}
respectively, where $\mu_B$, $\mu_Q$ and $\mu_I$ are the baryon, charged, and isospin chemical potentials. $Y_I$ is the isospin fraction defined as
\begin{eqnarray}
Y_I = \frac{I}{B} = \frac{\sum_i Q_{I,i} \ n_i} {\sum_i Q_{B,i} \ n_i} .
\label{6h}
\end{eqnarray}

Note that the free energy is equal to the baryon chemical potential only in the particular cases of zero electric charge or zero charged/isospin chemical potential. This is the case for deleptonized cold neutron stars (charge neutral in weak equilibrium $Y_Q=0$) and relativistic HICs ($Y_I=0$ with $\mu_I=0$ ). Eq.~\ref{ve2Q} was derived and discussed in detail in the Appendix D of Ref.~\cite{Hempel:2013tfa}, Eqs.~\ref{ve2Q} and \ref{ve2} were derived with an extra term for the general case in which net strangeness is constrained in Ref.~\cite{Aryal:2020ocm}, and  Eq.~\ref{ve2Q} was derived with an extra term for the case with trapped neutrinos and fixed lepton fraction in Ref.~\cite{Roark:2018uls}.

\begin{figure*}[t!]
\centering
\begin{minipage}{17.96pc}
\includegraphics[width=\textwidth]{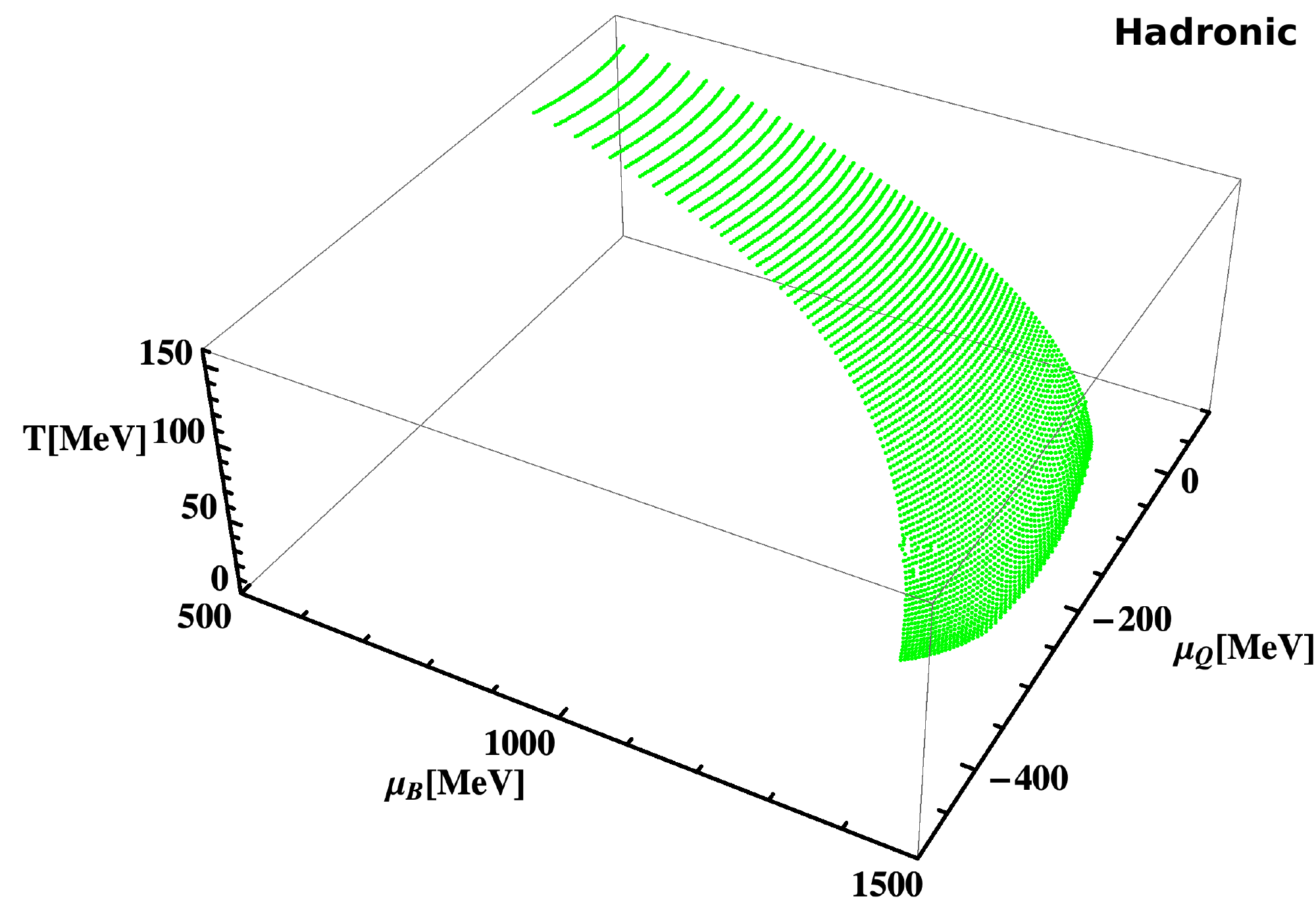}
\end{minipage}\hspace{2pc}%
\begin{minipage}{17.96pc}
 \includegraphics[width=\textwidth]{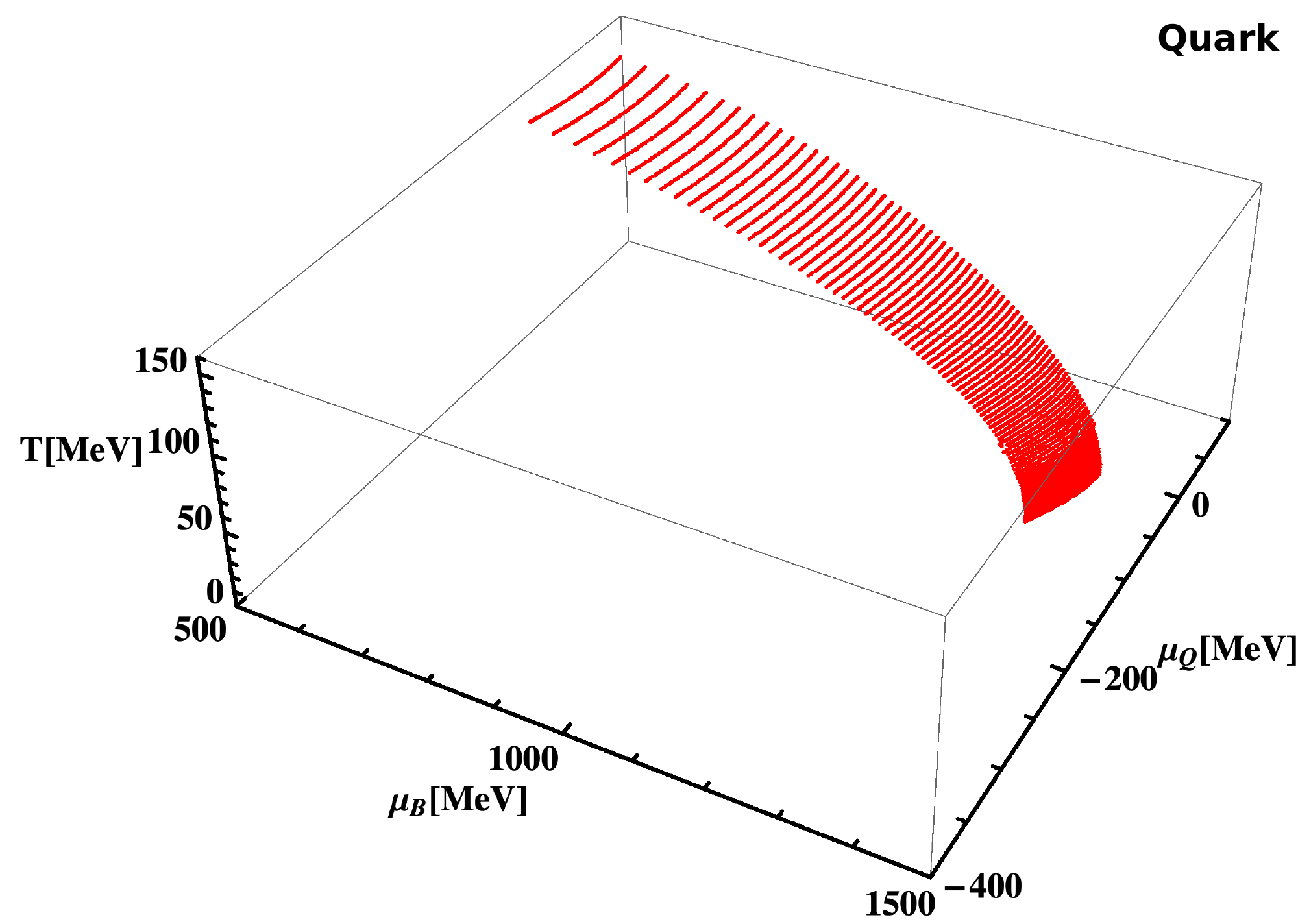}
\end{minipage}\hspace{2pc}%
\begin{minipage}{17.96pc}
\includegraphics[width=\textwidth]{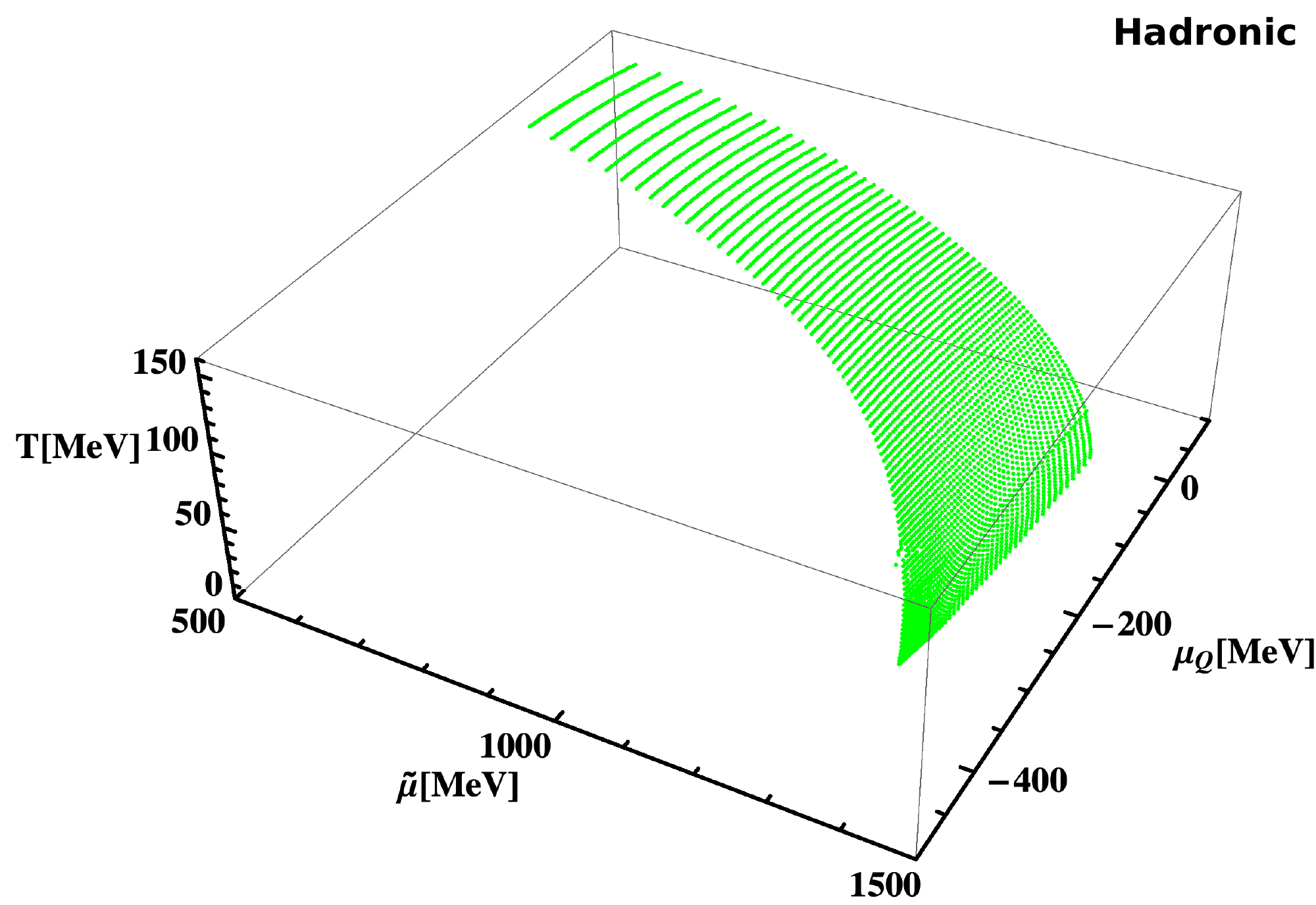}
\end{minipage}\hspace{2pc}%
\begin{minipage}{17.96pc}
\includegraphics[width=\textwidth]{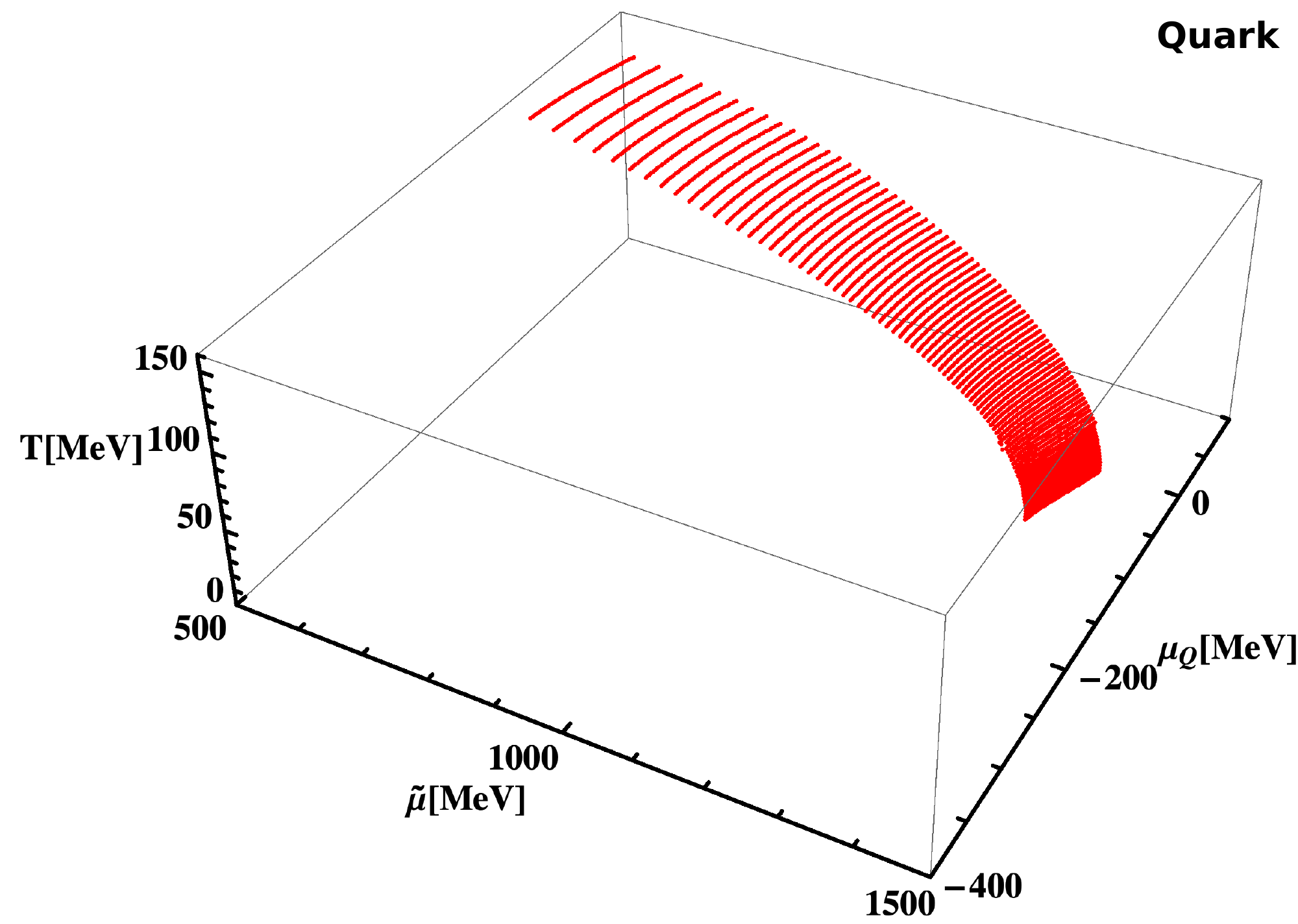}
\end{minipage} 
   \caption {Same as Fig.~\ref{fig:fig} but showing the charged chemical potential $\mu_Q$. The separate bottom panels show the hadronic (left panel) and quark (right panel) sides of the deconfinement phase transition}
\label{fig2:fig}
\end{figure*} 

The bottom panel of Fig.~1 shows how the free energy at the deconfinement coexistence line increases as a function of $Y_Q$. This behavior is related to the softening of nuclear matter with increased net charge (larger proton-to-neutron or up-to-down quark ratio), the effect being stronger for hadronic matter. A softening of the equation of state (pressure vs. energy density) of hadronic matter corresponds to an increase in pressure at a given free energy (with respect to the quark phase)  and, therefore, the hadronic phase persists to larger free energies. 

The difference between the top panels of Fig.~1 comes from the fact that the baryon chemical potential is calculated from the free energy using Eq.~\ref{ve2Q} with a different charged chemical potential on either side of the phase transition. When comparing the top left panel with the bottom one, there is a significant difference for all cases corresponding to $\mu_B \neq \widetilde{\mu}$ (for all $Y_Q$ other than 0 and 0.5, the latter implying $\mu_Q=0$). Note that the difference is much smaller between the top right panel of Fig.~1 and the bottom one, as the charged chemical potential $\mu_Q$ is always small (in absolute value) in the quark phase (see Fig.~3 of Ref.~\cite{Roark:2018uls} for the particular case of chemically equilibrated matter).

The charged chemical potentials are shown in Fig.~2 on both sides of the coexistence line, reaching positive values at large temperatures and more negative values in the hadronic phase (left panels) in comparison to the quark phase (right panels). A larger (in absolute value) charged chemical potential means a larger difference between, for example, the number density of protons and neutrons or up and down quarks. In Fig.~2, the bottom panels are always different from each other, since the charged chemical potential itself is discontinuous across the first-order phase transition.

In order to draw 3-dimensional phase diagrams using isospin axes, one either rewrites the formalism using isospin as a conserved quantity (and corresponding computer codes), or uses the following transformation
\begin{eqnarray}
Y_I =Y_Q - \frac{1}{2} +\frac{1}{2}Y_S ,
\label{vYi} 
\end{eqnarray} 
which was derived in detail in Ref.~\cite{Aryal:2020ocm}. The strangeness fraction $Y_S$ is defined as the other fractions but now accounting for the strangeness of each hadron. In our notation, we consider the strangeness of particles to be positive, otherwise, all strangeness related quantities would have their signs reversed. Eq.~\ref{vYi} implies that for non-strange matter ($Y_S=0$),  the change of variable from $Y_Q$ to $Y_I$ introduces only a simple shift of the phase diagram relative to that axis, $Y_I=Y_Q-0.5$. When net strangeness is finite ($Y_S \ne 0$), and to an increasing extent as temperature is raised, phase diagrams drawn against an isospin axis receive nontrivial modifications allowing the deconfinement coexistence line to cross to positive $Y_I$ values for large charge fractions $Y_Q \simeq 0.5$. This can be seen in the green, red, and brown regions of Fig.~3.

Additionally, all panels of Fig.~3 display blue regions. These were calculated by rewriting our computer code in terms of the isospin fraction $Y_I$, which runs from $-0.5$ to 0 (instead of varying the charge fraction in the range $Y_Q=0\to0.5$). As a result, these blue regions cover a different part of the QCD phase diagram. Note that, as $Y_S \rightarrow 0$ for low temperatures, the blue regions are deformed, such that they become identical with the other ones.

\begin{figure*}[t!]
\centering
\begin{minipage}{18.83pc}
\includegraphics[width=\textwidth]{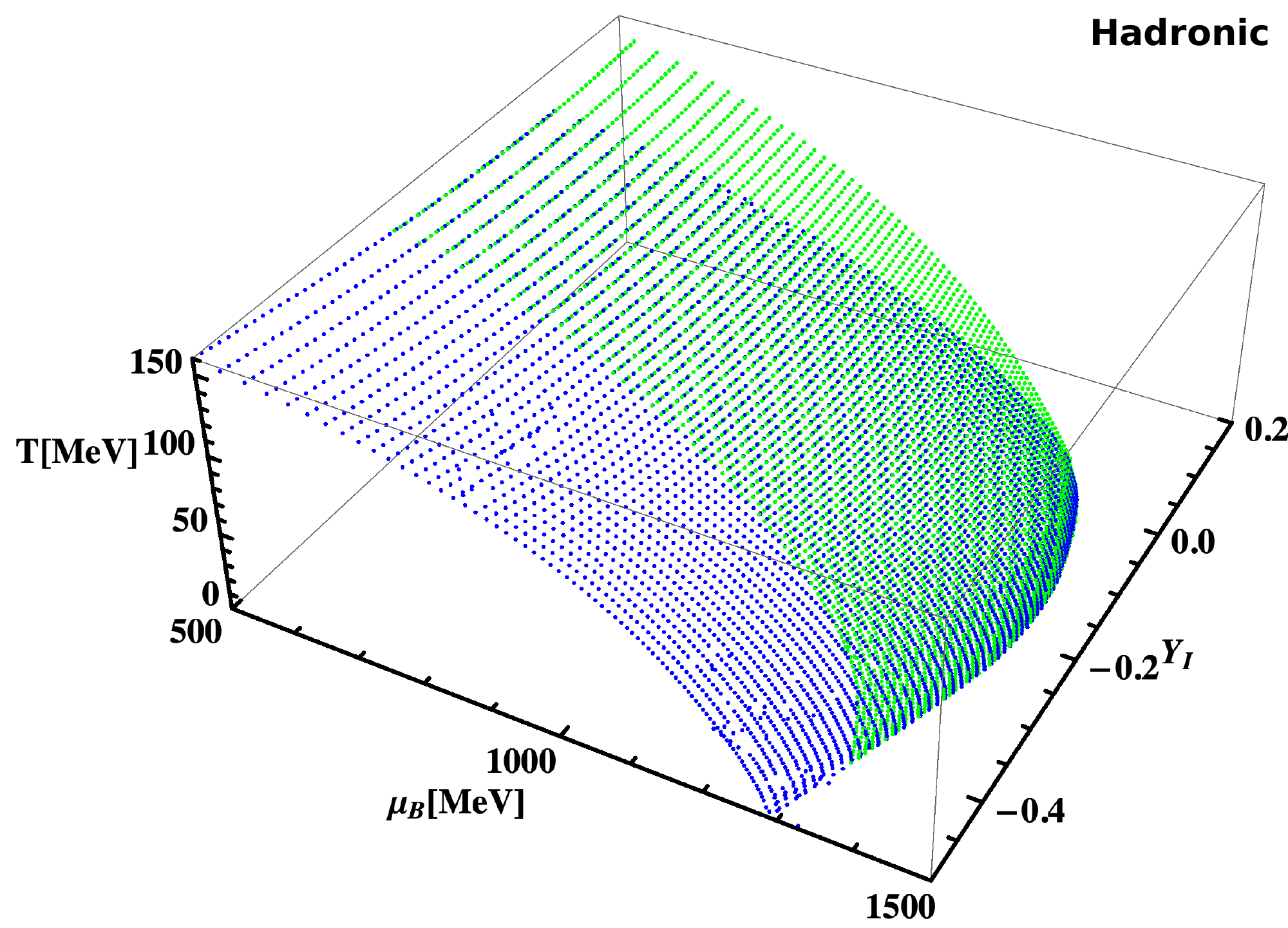}
\end{minipage} 
\begin{minipage}{18.83pc}
 \includegraphics[width=\textwidth]{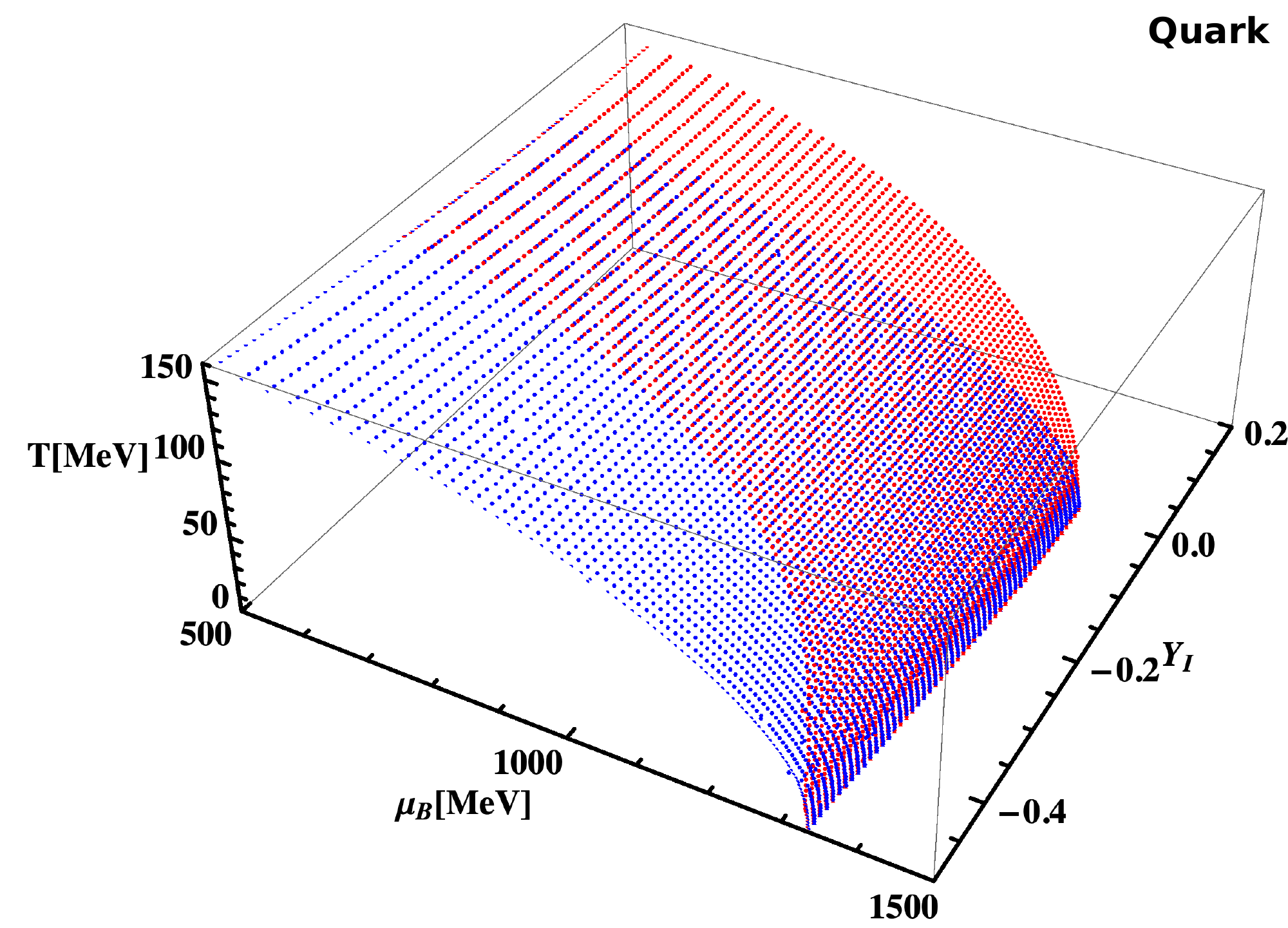}
\end{minipage} 
\begin{minipage}{18.83pc}
\includegraphics[width=\textwidth]{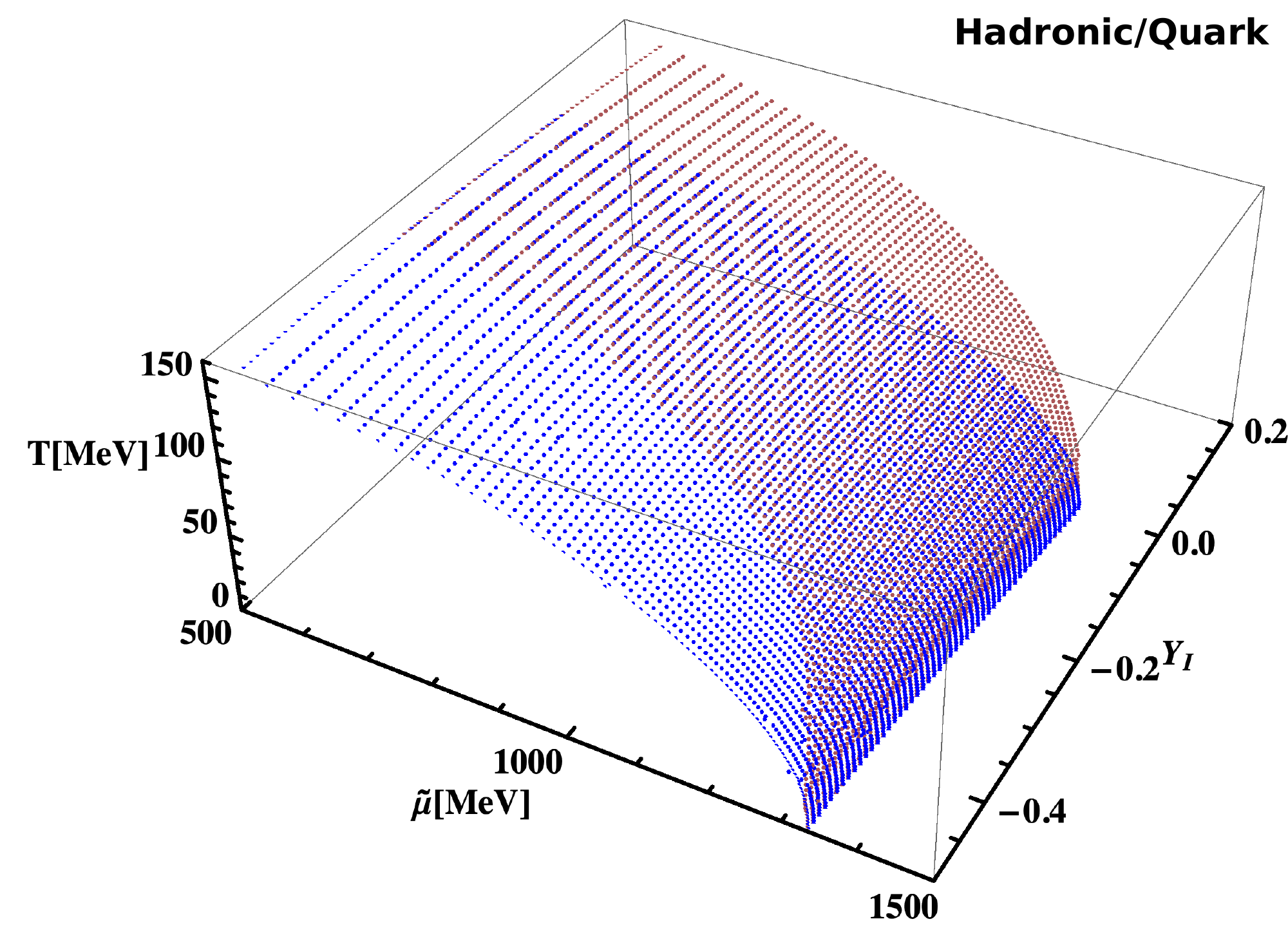}
\end{minipage} 
  \caption {Same as Fig.~\ref{fig:fig} but showing the isospin charge fraction $Y_I$. Additional blue regions show results obtained by varying the isospin fraction in the range $Y_I=-0.5\to0$ (instead of varying the charge fraction in the range $Y_Q=0\to0.5$).}  
\label{fig3:fig}
\end{figure*}

As can be seen in Fig.~4 (when compared to Fig.~2), both charged and isospin chemical potentials are equal. This is the case when using the following (non-unique) definitions based on the conserved quantities baryon and charge or isospin to calculate the chemical potential for each particle
\begin{eqnarray}
\mu_i & = & Q_{B,i} \ \mu_B + Q_i \ \mu_Q , 
\label{equation:mui}
\end{eqnarray}
or
\begin{eqnarray}
\mu_i & = & Q_{B,i} \ \mu_B + (Q_{I,i} + {{1}/{2}Q_{B,i}} -{1}/{2}Q_{S,i}){\mu_I} ,
\label{equation:mui2}
\end{eqnarray}
where $Q_I$ and $Q_S$ refer to the isospin and strangeness of each hadron or quark. Note that, even in the case that strangeness is not conserved, Eq.~\ref{equation:mui2} (derived in detail in Ref.~\cite{Aryal:2020ocm}) still contains a strangeness related term. The extra terms in the parentheses  are required to obtain the same standard chemical equilibrium equations for all hadrons and quarks (shown in Appendix A of Ref.~\cite{Aryal:2020ocm}), independently of the chosen formulation. As in Fig.~3, all panels of Fig.~4 display blue regions calculated by rewriting our computer code in terms of the isospin fraction $Y_I$ (instead of $Y_Q$). Within the blue regions, the isospin chemical potential $\mu_I$ remains negative.

\section{Final Remarks}

\begin{figure*}[t!]
\centering
\begin{minipage}{18.83pc}
\includegraphics[width=\textwidth]{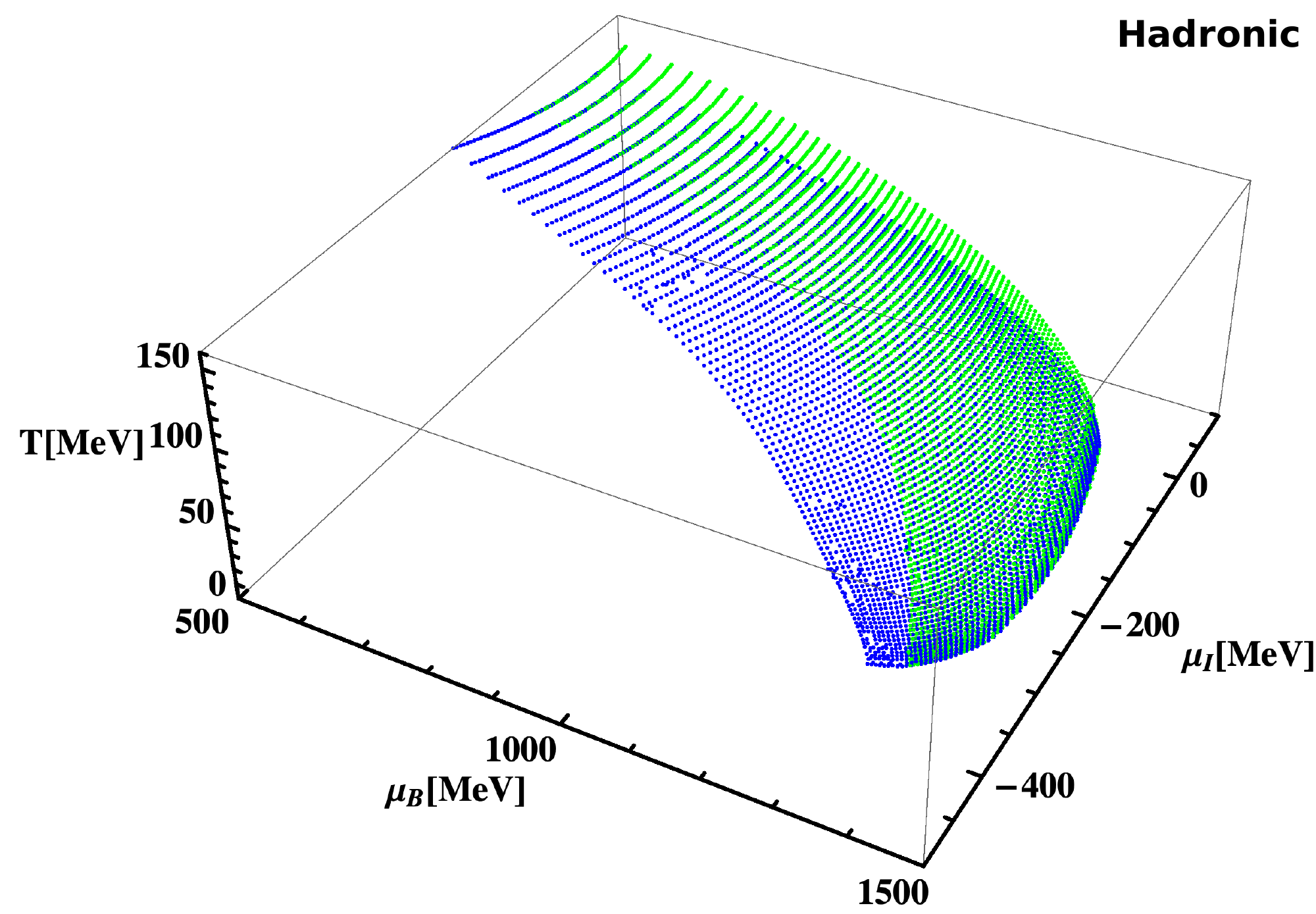}
 \end{minipage} 
\begin{minipage}{18.83pc}
 \includegraphics[width=\textwidth]{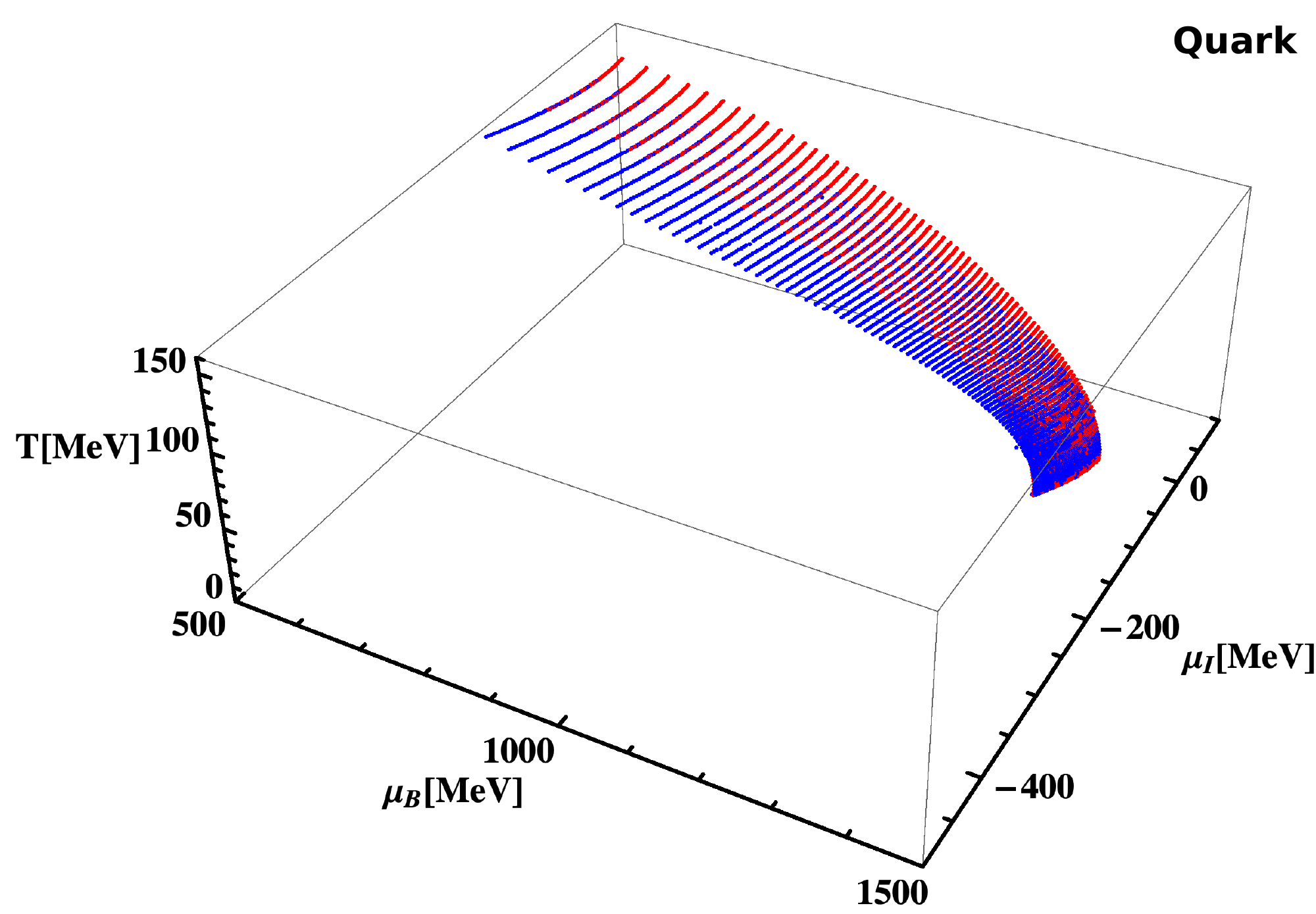}
\end{minipage} 
\begin{minipage}{18.83pc}
\includegraphics[width=\textwidth]{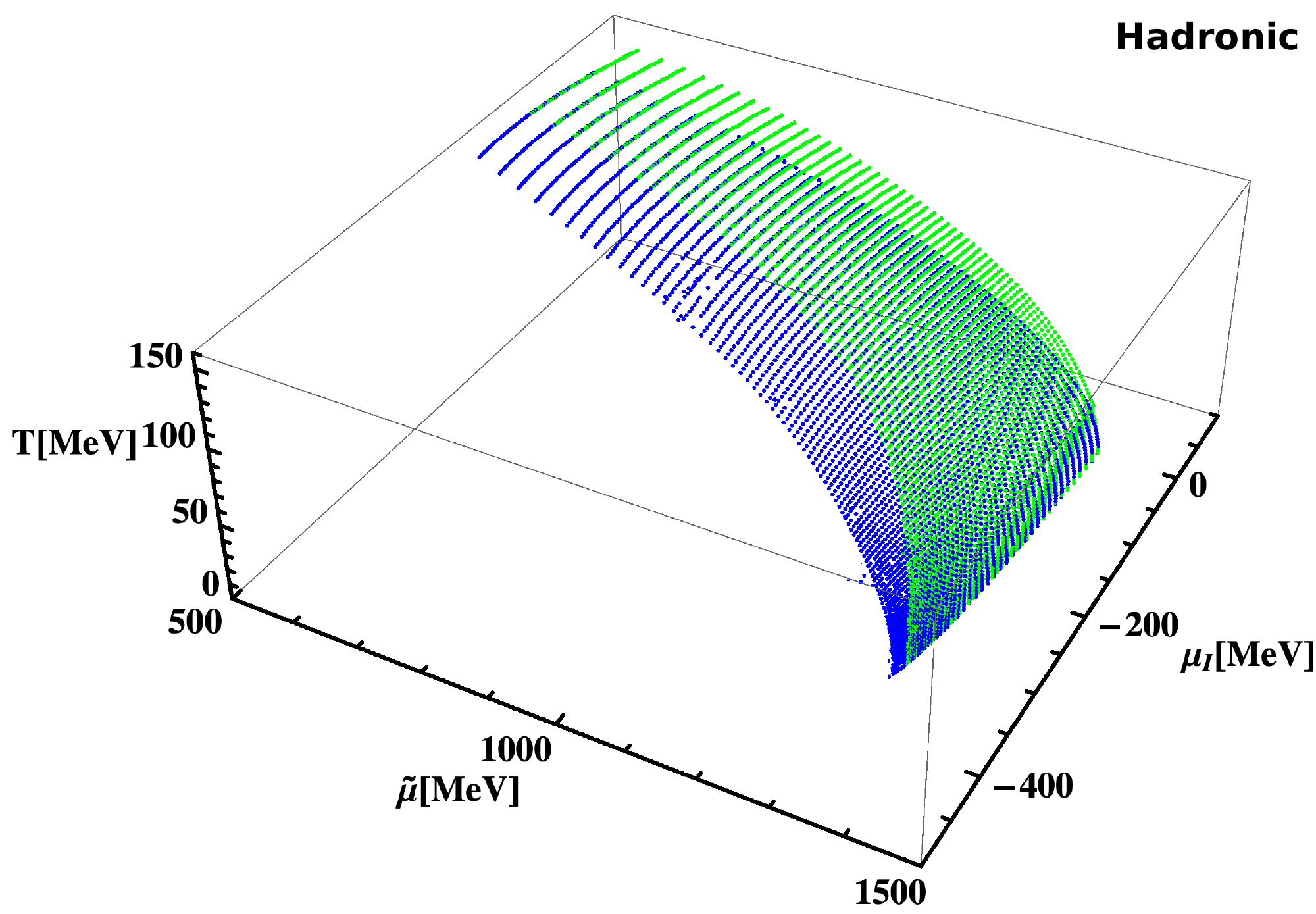}
\end{minipage} 
\begin{minipage}{18.83pc}
\includegraphics[width=\textwidth]{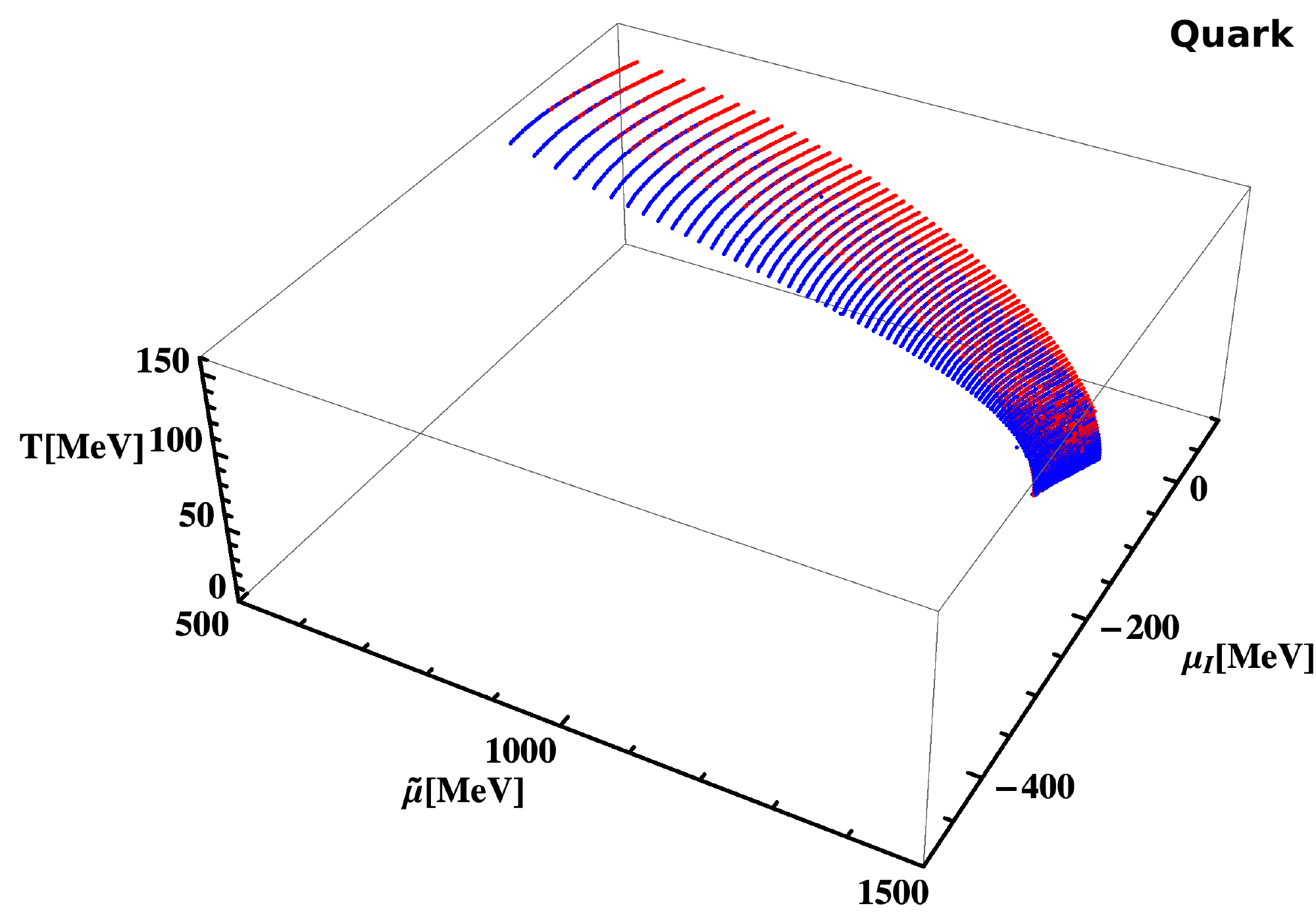}
\end{minipage} 
  \caption {Same as Fig.~\ref{fig2:fig} but showing the isospin chemical potential $\mu_I$. Additional blue regions show again results obtained by varying the isospin fraction in the range $Y_I=-0.5\to0$ (instead of varying the charge fraction in the range $Y_Q=0\to0.5$).}   
\label{fig4:fig}
\end{figure*}  

In this work, we presented 3-dimensional QCD phase diagrams with the extra axis being the charge fraction/chemical potential or the isospin fraction/chemical potential.  We discussed these two different approaches for conserved quantities following different common practices in the astrophysics and heavy-ion collision communities. We found that changing the charge or isospin fraction by $0.5$ can change the position of deconfinement phase transition significantly. The baryon chemical potential can change up to $130$ MeV (at zero temperature on the hadronic side), the free energy up to $50$ MeV (at zero temperature), and the charge/isospin chemical potential by $330$ MeV (at zero temperature on the hadronic side). Our results show that comparisons among results from heavy-ion collision and hot astrophysical scenarios concerning the position of the deconfinement phase transition have to be interpreted carefully, as their different characteristics can change considerably the position of the phase-transition coexistence line.

\section*{Acknowledgements}
\vspace{1mm}

Support for this research comes from the National Science Foundation under grant PHY-1748621, PHAROS (COST Action CA16214), the LOEWE-Program in HIC for FAIR, Conselho Nacional de Desenvolvimento Cient\'{\i}fico e Tecnol\'ogico - CNPq under grant 304758/2017-5 (R.L.S.F), and Funda\c{c}\~ao de Amparo \`a Pesquisa do Estado do Rio Grande do Sul - FAPERGS under grants 19/2551-0000690-0 and 19/2551-0001948-3 (R.L.S.F.).

\section*{References}
\vspace{1mm}

\bibliography{paper}

\end{document}